\documentclass[a4paper,11pt]{article}
\pdfoutput=1
\usepackage{jinstpub}
\usepackage{lineno}
\usepackage{verbatim}

\title{\boldmath Silicon Carbide Timepix3 detector for quantum-imaging detection and spectral tracking of charged particles in wide range of energy and field-of-view}

\author[a,1]{A. Novak\note{Corresponding author.},}
\author[b]{C. Granja,}
\author[a]{A. Sagatova,}
\author[b]{J. Jakubek,}
\author[c]{B. Zatko,}
\author[d]{V. Vondracek,}
\author[d]{M. Andrlik,}
\author[e]{V. Zach,}
\author[b]{S. Polansky,}
\author[b]{A. Rathi,}
\author[b]{and C. Oancea,}

\affiliation[a]{Slovak University of Technology in Bratislava, Institute of Nuclear and Physical Engineering, \\Ilkovicova 3, 841 04, Bratislava, Slovakia}
\affiliation[b]{Advacam, U Pergamenky 12, 170 00, Prague 7, Czech Republic}
\affiliation[c]{Institute of Electrical Engineering, Slovak Academy of Sciences, 841 04 Bratislava, Slovakia}
\affiliation[d]{Proton Therapy Center Czech, 180 00 Prague 8, Czech Republic}
\affiliation[e]{Nuclear Physics Institute, Czech Academy of Sciences, 250 68, Rez near Prague, Czech Republic}

\emailAdd{andrej.novak@stuba.sk}

\abstract{
The hybrid architecture of the Timepix (TPX) family of detectors enables the use of different semiconductor sensors, most commonly silicon (Si), as well as high-density materials such as Cadmium Telluride (CdTe) or Gallium Arsenide (GaAs). For this purpose, we explore the potential of a silicon carbide (SiC) sensor bump-bonded on a Timepix3 detector as a radiation imaging and particle tracking detector. SiC stands as a radiation-hard material also with the ability to operate at elevated temperatures up to several hundreds of degrees Celsius. As a result, this sensor material is more suitable for radiation harsh environments compared to conventional e.g., Si sensors. In this work, we evaluate the response for precise radiation spectrometry and high-resolution particle tracking of newly developed SiC Timepix3 detector which is built and operated as a compact radiation camera MiniPIX-Timepix3 with integrated readout electronics. Calibration measurements were conducted with mono-energetic proton beams with energies of 13, 22, and 31 MeV at the U-120M cyclotron at the Nuclear Physics Institute Czech Academy of Science (NPI CAS), Prague, as well as 100 and 226 MeV at the Proton Therapy Center Czech (PTC) in Prague. High-resolution pattern recognition analysis and single-particle spectral tracking are used for detailed inspection and understanding of the sensor response. Results include distributions of deposited energy and linear energy transfer (LET) spectra. The spatial uniformity of the pixelated detector response is examined in terms of homogeneously distributed deposited energy.}

\keywords{Particle tracking; Silicon-Carbide semiconductor sensor; Pixel detector Timepix3; Radiation imaging}

\begin{document}
\maketitle
\flushbottom

\section{Introduction}
\label{sec:intro}

Timepix detectors are used in a broad range of applications~\cite{HEI13,BAL18} ranging from medical imaging, material research to space radiation monitoring~\cite{single_LET, proton_pencil_beam, imaging_1, imaging_2, radiation_monitor}. These environments are characterised by high-intensity radiation fields that degrade and damage the detectors exposed for an extended period of time. 
Currently, commercially available sensors for Timepix radiation detectors are made of Si and CdTe. However, in response to the challenge of radiation-induced damage, ongoing investigations are exploring alternative, radiation-hard materials, including GaAs~\cite{gaas}, diamond~\cite{diamond}, or SiC~\cite{b}. The SiC sensor with a 3.23 eV band gap also offers a high breakdown voltage ($4\times10^6$ V m${}^{-1}$), which allows the application of a large bias voltage that leads to fast charge collection. It has been demonstrated that the energy resolution performance of SiC is not degraded even at elevated temperatures up to hundreds of degrees Celsius~\cite{sic_temperature}. Furthermore, the material itself has high electron mobility (around 900 cm${}^2$ V${}^{-1}$ s${}^{-1}$) and high electron saturation drift velocity ($2\times 10^7$ cm${}^{-1}$). The primary significance of SiC lies in its application within radiation-challenging environments, requiring detectors that can endure prolonged radiation exposure. Potential uses include high-energy physics, outer space and particle radiotherapy. 
For this purpose, a fully operational miniaturized MiniPIX Timepix3 radiation camera, featuring an 80 $\mu$m thick SiC epitaxial layer, has been developed and built including energy calibration and SW/FW configuration. 
In this work we characterise this novel SiC Timepix3 detector in terms of its homogeneity, charge collection response, and spectral tracking response with proton beams of energy range 13 to 226 MeV. This comprehensive analysis aims to facilitate the use of a radiation hard sensor in various applications such as space radiation monitoring, neutron detection, or conventional and FLASH radiotherapy environments~\cite{flash_stray}.

The Timepix detectors provide valuable capabilities of position-, time- and directional sensitivity which make them suitable for selective detection and high-resolution wide-range spectral tracking of single particles~\cite{GRA18,tpx2_mixed_field}. The detector high granularity and per-pixel spectral response can be used to examine and evaluate the uniformity and homogeneity of the radiation response and charge collection of the semiconductor sensor~\cite{novak}.

\section{Instrumentation and experiments}

As part of our investigation into the behaviour of SiC material, we performed a comprehensive characterisation of a Schottky 4H-SiC single-pad detector. This step served as a foundation in the development of the pixel structure for the hybrid semiconductor pixel detector Timepix3 with a SiC sensor. The alpha particles generated by the Am-241 radioisotope and detected by the SiC Schottky detector were resolved with a full width at half maximum (FWHM) of 13.8 keV at 5480 keV peak at room temperature. The single-pad Schottky barrier SiC detectors were developed to investigate and optimise their electrical and detection properties. The evaluation of 4H-SiC Schottky barrier detector performance at elevated temperatures is presented in~\cite{sic_temperature}, where we demonstrated the detector’s operation up to 500$^{\circ}$C. At this high-temperature, the FWHM degraded $\sim$20\% when compared to its performance at room temperature. The 4H-SiC detectors have been proven to be suitable for high-resolution spectrometry of alpha particles in a wide range of elevated temperatures.

\subsection{Radiation camera Minipix Timepix3 with SiC sensor}

\begin{figure}[htbp] 
\centering
\includegraphics[width=0.35\textwidth]{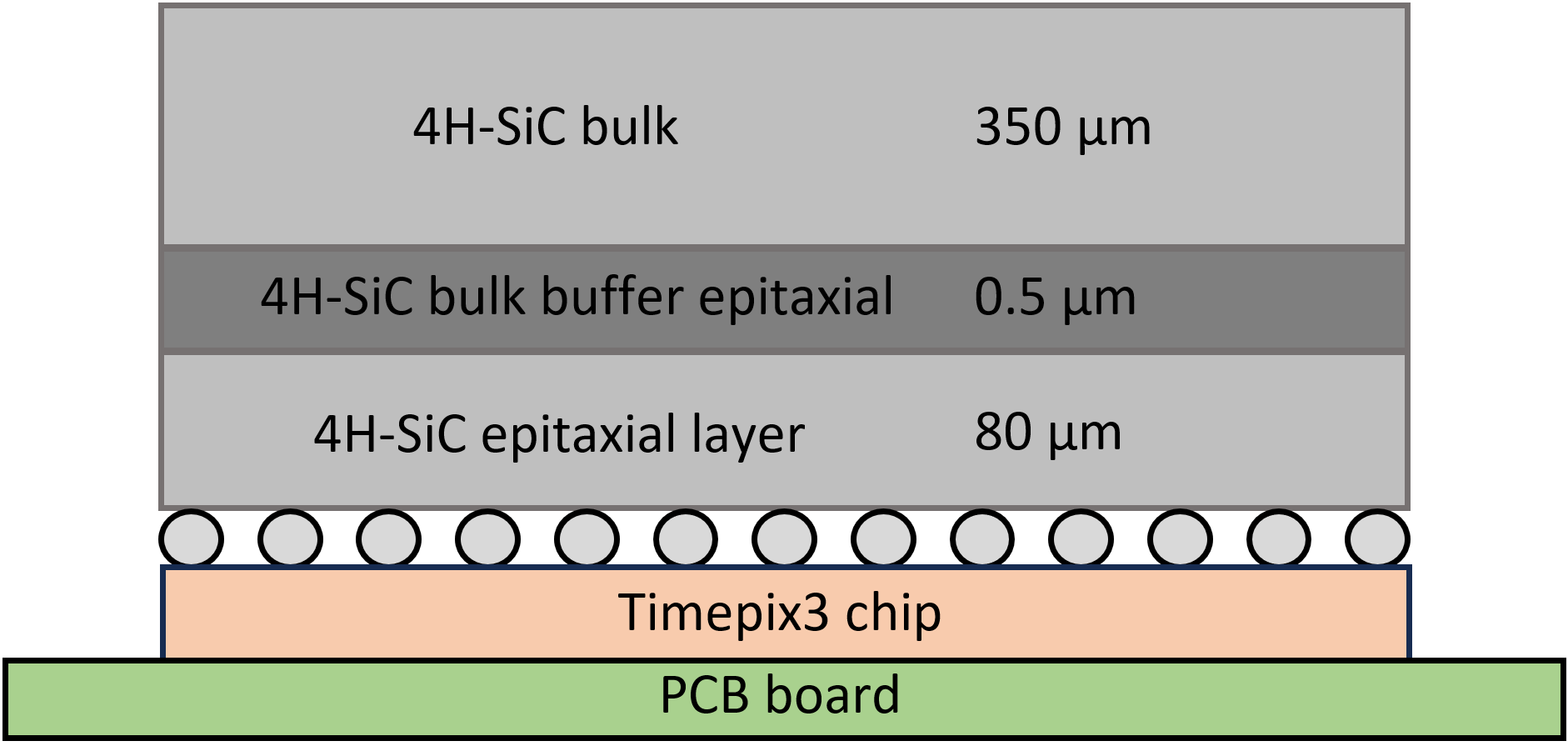}
\quad
\includegraphics[width=0.30\textwidth]{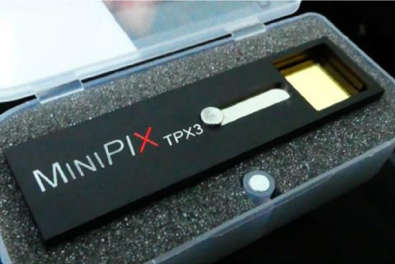}
\caption{\label{fig:diagram_and_minipix} Left: Diagram of the cross section of the SiC sensor showing the bulk, buffer, and epitaxial grown layers. The latter contains, by full depletion +200V applied bias voltage, the radiation sensitive volume 65~$\mu$m thick facing down to the Timepix3 ASIC bump contacts. The remaining volume 365.5 $\mu$m thick facing up is regarded as a nonsensitive absorption layer. Right: The compact radiation camera Timepix3 MiniPIX equipped with a SiC sensor operated with a single USB connector.}
\end{figure}

Based on the 4H-SiC epitaxial layer characterisation demonstrated in single-pad detectors, the pixel structure was prepared for the hybrid pixel detector Timepix3. For the investigation of SiC as a sensor material for the Timepix3 chip, two companies prepared four 4H-SiC sensors. Epitaxial growth of SiC was used to achieve 80 $\mu$m and 100 $\mu$m thick layer, respectively. After bump-bonding of the sensors to the Timepix3 chip, the detectors were connected to the MiniPIX readout; see Fig.~\ref{fig:diagram_and_minipix} (right). In this study, the sensors used were limited to a bias of 200 V to prevent electrical shortages. However, due to the limited bias voltage, only 65 $\mu$m out of 80 $\mu$m detector sensor thickness could be fully depleted. On top of the 80 $\mu$m active layer (nitrogen doped layer, doping concentration 1$\times$10${}^{14}$ cm${}^{-3}$) there is a buffer (concentration 1$\times$10${}^{18}$ cm${}^{-3}$) and a bulk layer (2$\times$10${}^{18}$ cm${}^{-3}$) which together with the non-depleted epitaxial region form a 365.5 $\mu$m thick dead layer that absorbs radiation; see Fig.~\ref{fig:diagram_and_minipix} (left).

As a result, low-energy X-ray imaging requires extended exposure time in order to acquire a significant number of counts. During the process of standard detector energy calibration using X-ray fluorescence, the exposure time has been adjusted to address these limitations. The description of the calibration process for Minipix Timepix detectors is described in~\cite{jakubek}. Alpha particle measurements using laboratory isotopes such as Am-241 (5.5 MeV) were not feasible, given the particle range is in order of microns and stops within the detector material.  The sensor is suitable for the detection of accelerated high-energy light and heavy ions that can transverse the active volume. An alternative approach for particle tracking and detection of accelerated particles is to tilt the sensor to enable the particle beam to impinge on the sensor from the side.

\subsection{Measurements with 13, 22, 31, 100 and 226 MeV protons}

In this work, we evaluate the response of the SiC Timepix3 detector on monoenergetic protons of various energies incident on a wide range of directions (from the perpendicular direction ($0^{o}$) to the parallel ($90^{o}$ direction). 
Low-energy protons (13, 22 and 31 MeV) were measured at the light ion cyclotron U-120M~\cite{KRI18} of the NPI CAS Rez near Prague. 
High-energy protons (100 and 226 MeV) were measured at the IBA Proteus-235 cyclotron at the Proton Therapy Center Prague. Measurements were performed in air, at 2 m and 80 cm distance from the accelerator beam nozzle. The pixel detector was rotated to the proton beam direction at various angles in the full angular range. By this approach it can be tuned the particle track length within the active region, resulting in a variable and well-defined energy deposition by charged particles at specific incident angles. 

\begin{figure}[htbp] 
\centering
\includegraphics[height=0.40\textwidth]{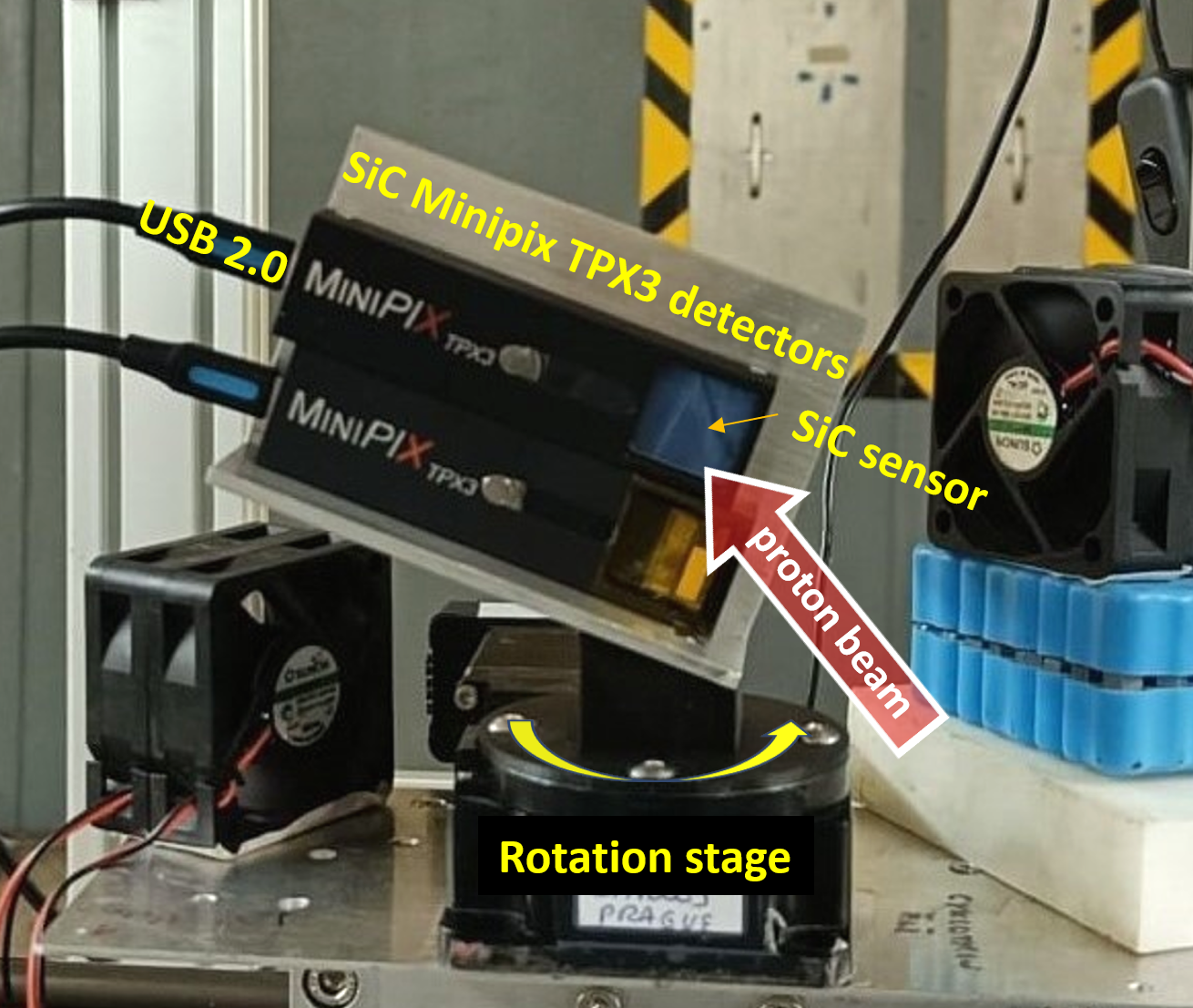}
\includegraphics[height=0.42\textwidth]{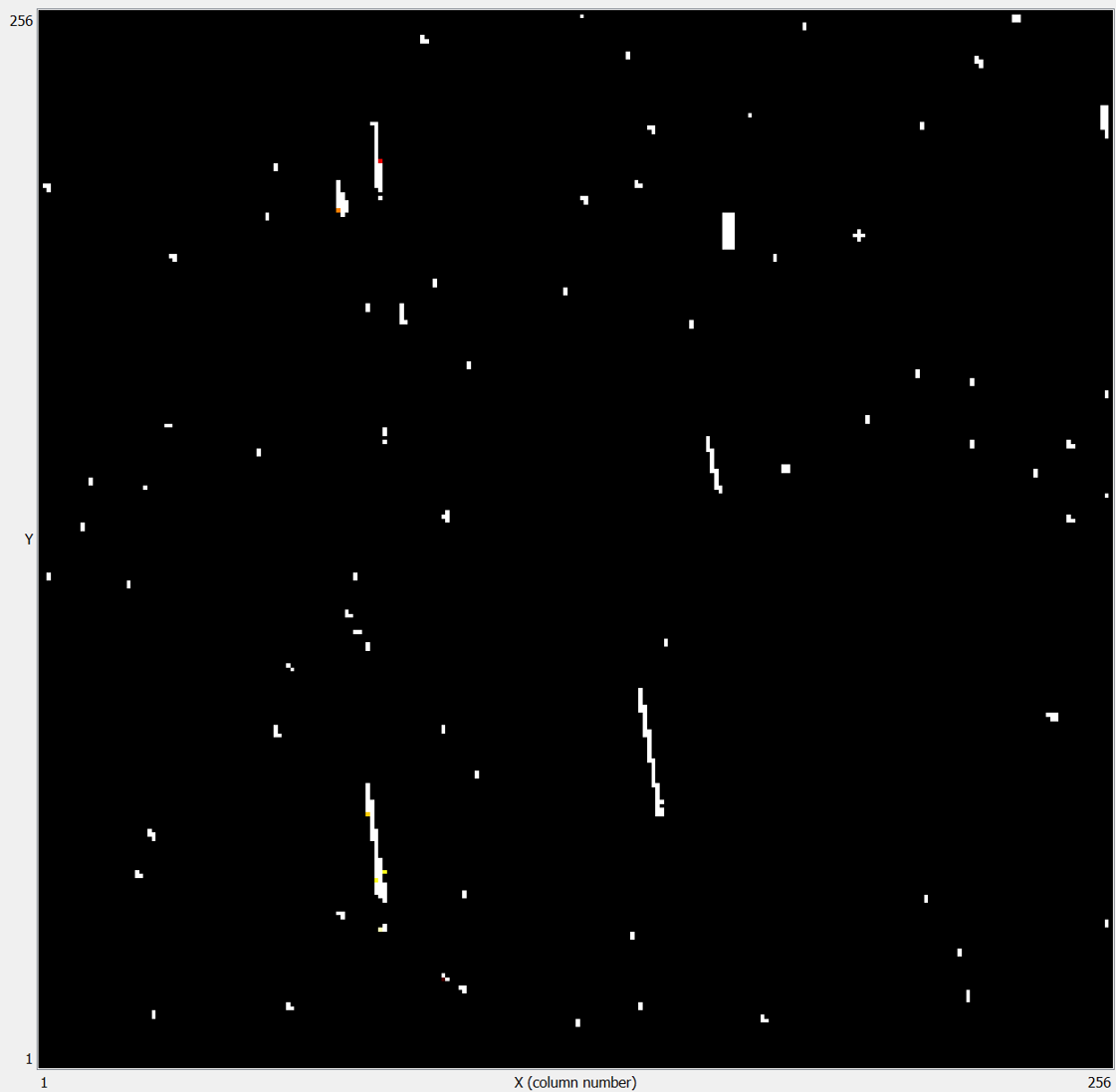}
\caption{\label{fig:rotation_stage_frame} 
Left: Experimental setup with two SiC Timepix3 MiniPIX detectors mounted on a rotation stage for proton beam measurements at the NPI-CAS Rez light-ion cyclotron. Rigth: Detection of 100 MeV protons incident at 88$^{\circ}$ (long tracks pointing down) by SiC Timepix3 with 65~$\mu$m thick active volume. Measured at PTC Prague. The full detector pixel matrix is shown (256 $\times$ 256 pixels = 14 mm $\times$ 14 mm). Unwanted/background tracks are also registered. Data displayed in counting mode.}
\end{figure}

\section{Spectral-sensitive tracking of protons} 
 
\subsection{Data processing}
The active region of the detector consists of a matrix with 256$\times$256 pixels, each with a pixel pitch of 55 $\mu$m, resulting in a total of over 65 thousand pixels. Each pixel functions as an individual semiconductor detector. When high-energy particles traverse the sensor, they ionise its volume, leading to charge collection and the generation of multi-pixel events. These individual pixels are grouped into clusters based on their spatial and temporal characteristics. 
High-resolution pattern recognition algorithms~\cite{HOL08} are then applied to analyse the single particle cluster tracks and resolve particle-type events~\cite{GRA18}. These parameters allows to select and examine specific types of particles in different radiation environments. Once we determined parameters that effectively described these distinct groups, we were able to filter and analyse them separately. These and further steps of data processing are made using the integrated data processing engine (DPE)~\cite{dpe}. Developed by ADVACAM, the DPE tool performs data clustering, applies per-pixel energy calibration, and calculates for each particle track morphological and spectral parameters, including cluster size, length, roundness, deposited energy and LET~\cite{CAR21LET,proton_pencil_beam, novel_LET}. 

\subsection{Particle-type discrimination}
Based on the parameters assigned to each cluster/particle detected, it is possible to process and filter each particle group in a mixed-field measurement separately~\cite{GRA18}. This approach effectively separates high-energy particles or other specific events of interest from the unwanted background, yielding clean datasets for specific particle types. The SiC sensor provides a limited active volume, buried beneath the non-sensitive bulk material above which acts as a shielding filter for the incident radiation. Therefore, only a small part of the energy deposited by passing high-energy particles is collected in the 65~$\mu$m thick volume of the SiC sensor. This limitation extends to other parameters, such as morphological characteristics, where the boundaries between different particle groups are less distinct than in thick (i.e., $\geq$~300~$\mu$m thick) fully depleted sensors such as Si, CdTe, or GaAs~\cite{novak}. Only a few parameters are suitable for high-energy particle filtration, and for this task, the deposited energy and cluster size have proven to be the most effective whilst maintaining the high quality of the cleanliness of the filtered dataset.

\subsection{Quantum-imaging detection} 

\begin{figure}[htbp] 
\centering
\includegraphics[width=0.99\textwidth]{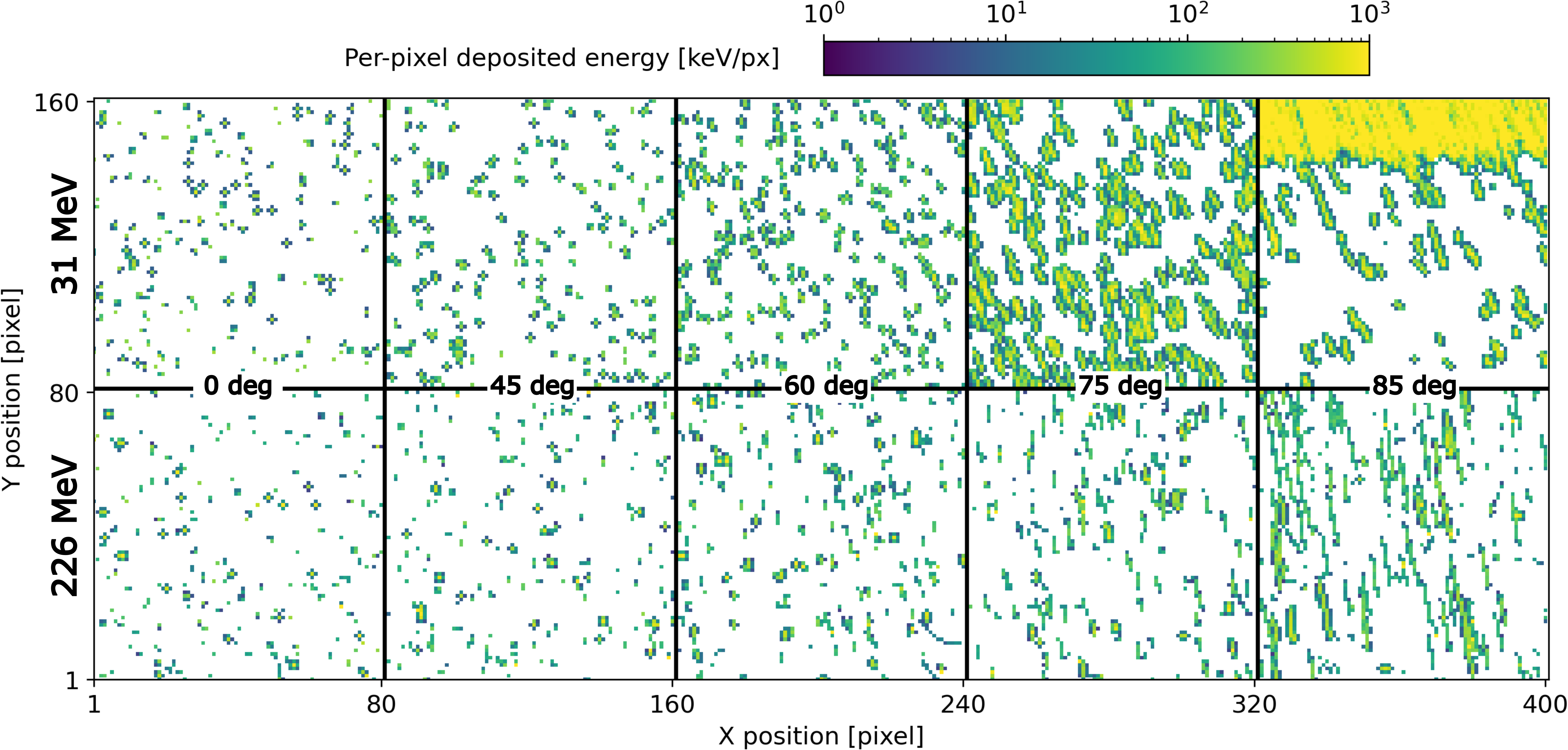}
\caption{\label{fig:LE_HE_rotation_calibration} Quantum-imaging detection and track visualisation of protons of 31 MeV (top row) and 226 MeV (bottom row) incident at varying direction with respect to the sensor plane. Only a part of the detector sensor area is shown - 80 px $\times$ 80 px = 4.4 mm $\times$ 4.4 mm = 0.19 cm${}^2$. The per-pixel energy registration is displayed by the color bar (log scale). Tracks are shown integrated over multiple frames. All particles are shown including background - e.g. small tracks by X-rays and electrons.}
\end{figure}

The spectral sensitive detection and track visualisation of 31 MeV and 226 MeV protons by SiC Timepix3 is shown in Fig.~\ref{fig:LE_HE_rotation_calibration}. 
In the figures, all registered particles are shown, no particle filter was applied. 
With increasing incident angle, the path of the proton track in the active sensor volume increases, leading to a higher deposited energy. 
For low-energy heavy charged particles incident at large angle, e.g., 31 MeV protons at $\geq 75^{\circ}$, the particle tracks approach the Bragg peak inside the sensor. For protons at 85$^{\circ}$ the Bragg peak is within the sensor active volume. For such low-energy protons, it can be observed the effect of shielding by the casing edge (2 mm Al) of the MiniPIX camera. 

\subsection{Deposited energy distributions} 
\begin{figure}[htbp] 
\centering
\includegraphics[width=0.98\textwidth]{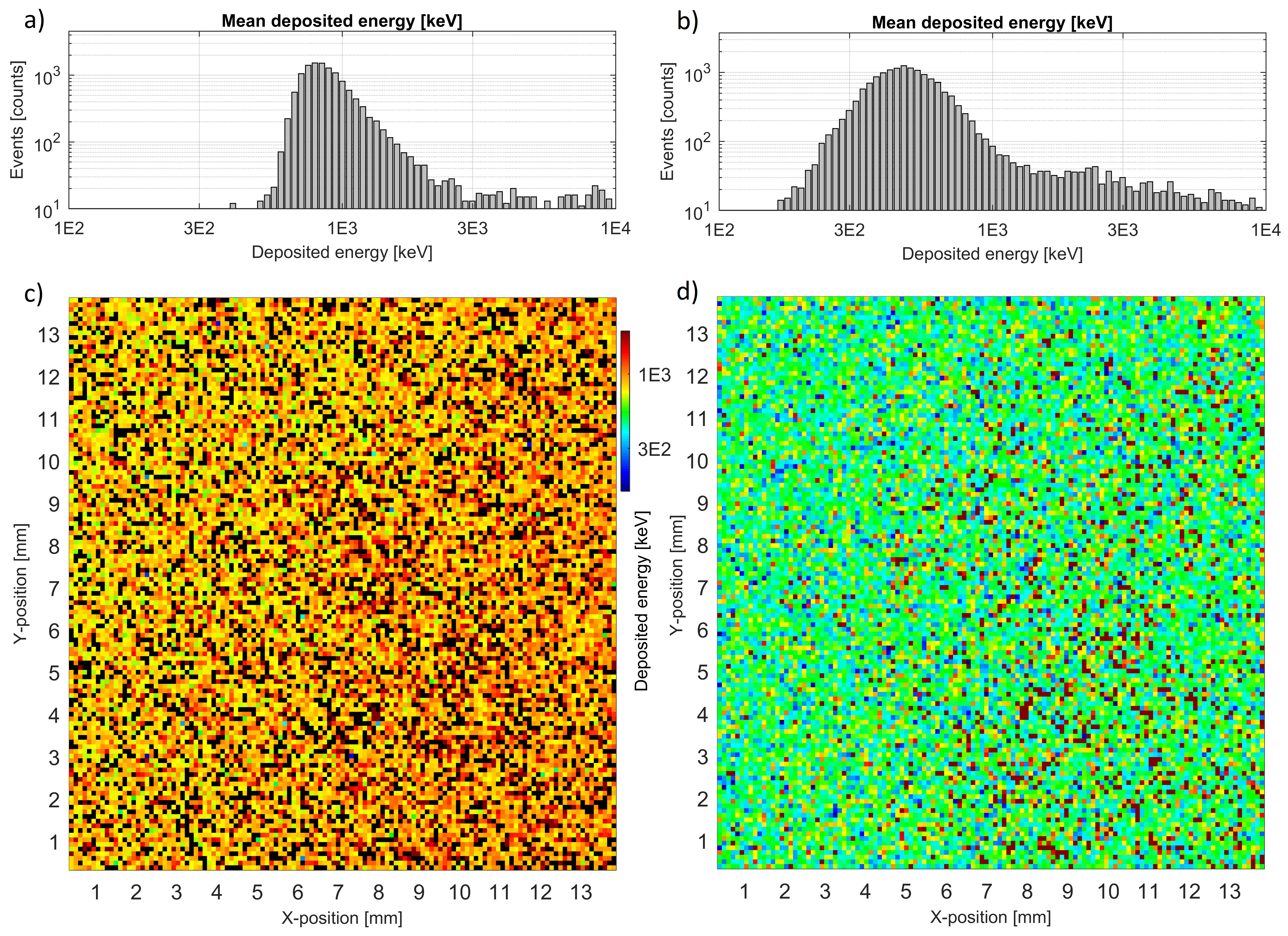}
\caption{\label{fig:spatial_homogeneity} 
Deposited energy in SiC Timepix3 of protons of 13 MeV (left) and 226 MeV (right) incident perpendicular, i.e., 0$^{\circ}$. The distributions are given as energy spectra (top row) and spatial map over the detector pixel matrix (bottom). Values presented for rebinned spatial bins of two pixel (i.e., 110~$\mu$m) size.}
\end{figure}

The spectral response of the Timepix3 detector can be used to measure in detail the deposited energy of single particles. This work also serves to examine and evaluate the spectral sensitivity, range, and homogeneity of the SiC sensor. The results for protons of selected energy (13 and 226 MeV) and direction (perpendicular 0$^{\circ}$) are given in Fig.~\ref{fig:spatial_homogeneity}. Background and unwanted events filtered (a filter pass was applied to the track cluster area of $\geq$ 3 pixels).

For the data evaluated, the mean values of the deposited energy are around 830 keV and 500 keV for 13 MeV and 226 MeV protons, respectively. 
Data exhibit the characteristic broad and right-asymmetric distributions of energy loss of charged particles in semiconductors~\cite{CAR21LET}. 
In addition, an anomalous large-amplitude component above a few MeV is observed. This is produced by events of high energy deposited up to nearly 10 MeV. These high-energy values are partly amplified by distortion and saturation of the per-pixel spectral response~\cite{SOM22}. 
Correction for this effect and examination of these large-amplitude events are the subject of future work. 

In Fig.~\ref{fig:spatial_homogeneity} it can be examined and evaluated the spatial homogeneity over the sensor-pixel matrix. 
The response map of the regular spectral component is evenly spatially distributed. The anomalous large-amplitude component exhibits a partly localised pattern which will be subject to further examination.

\subsection{LET spectra} 
\begin{figure}[htbp]
\centering
\includegraphics[width=0.48\textwidth]{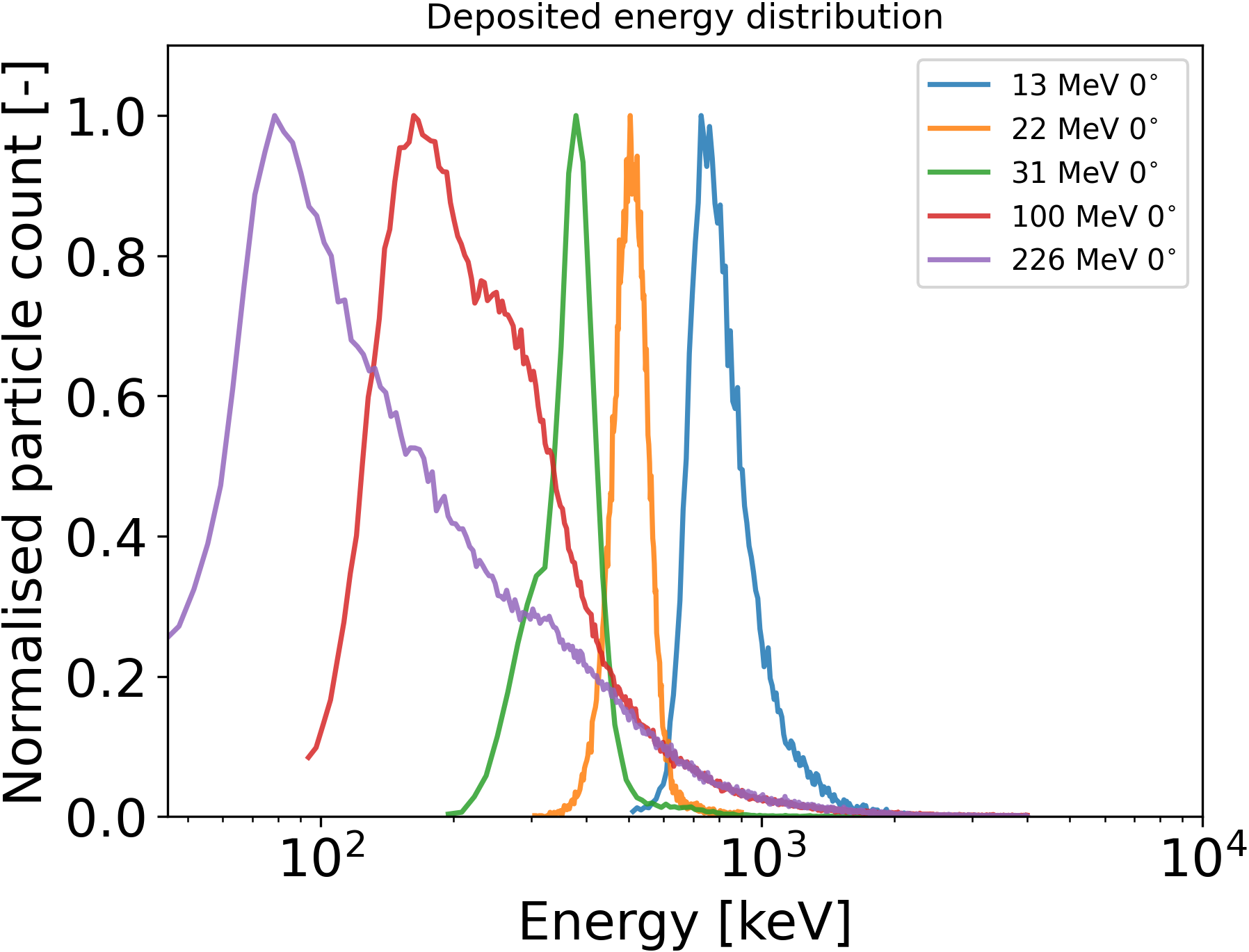}
\quad
\includegraphics[width=0.48\textwidth]{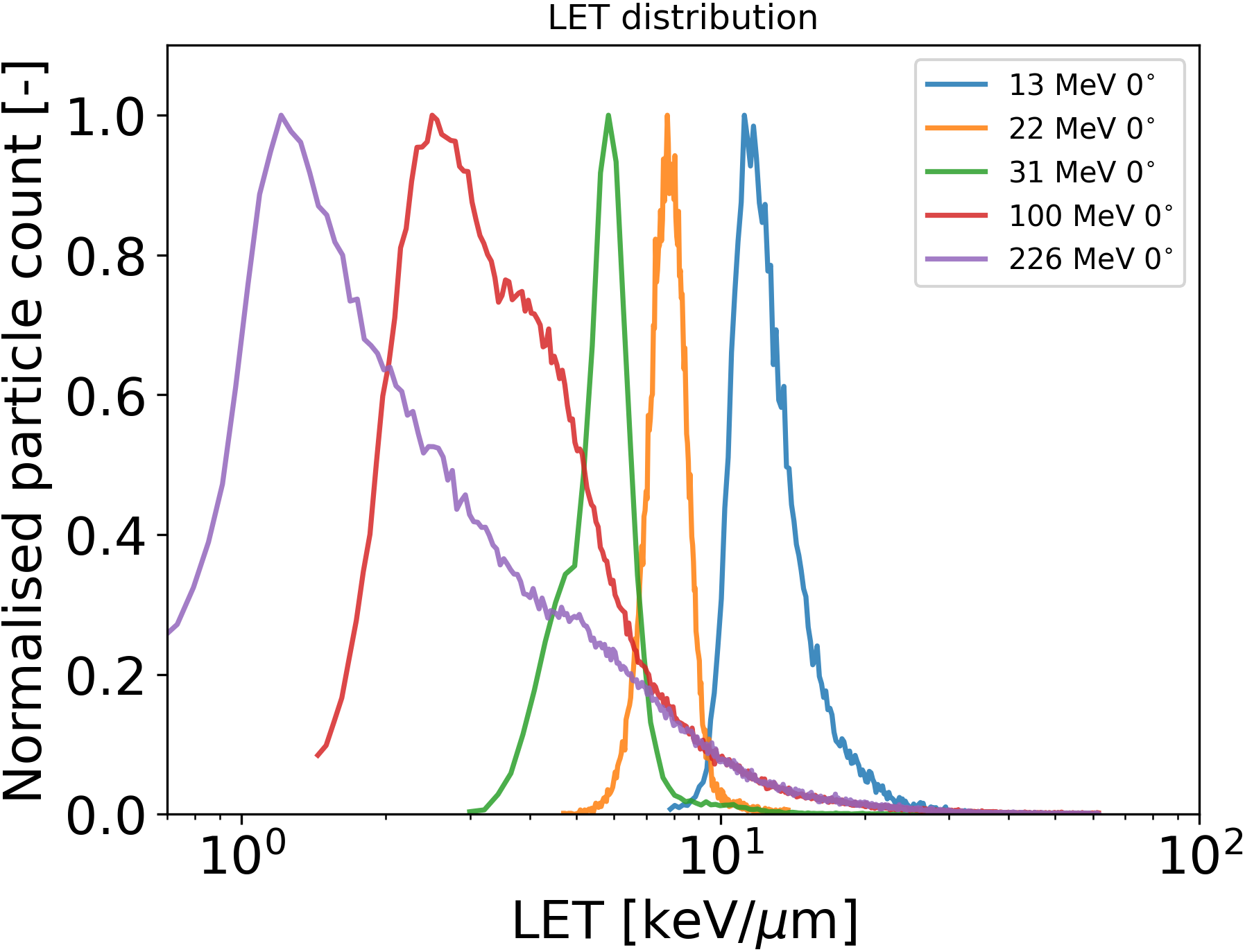}
\caption{\label{fig:distribution_E_len_LET} 
Deposited energy (left) and LET (right) spectra for protons of different energy. Measured by SiC Timepix3 placed perpendicular to the incident beam axis. A filter was applied to separate and analyse protons.}
\end{figure}

The high-resolution spectral-tracking response of Timepix3 enables to measure both the particle deposited energy and track length across the sensor active volume. This enables one to derive the LET of single particles in a wide range of energy and direction. The results of the selected data are given in Fig.~\ref{fig:distribution_E_len_LET} for protons of different energy incident perpendicularly to the sensor plane. 
For the SiC sensors used (see Fig.~\ref{fig:diagram_and_minipix}), the incident particles must first cross over the non-sensitive volume (365.5 $\mu$m thick inactive layer) before reaching the active volume with a thickness of 65 $\mu$m. For protons this effect results in a partly increased deposited energy and a right-shift of the peak position in the spectra. This effect can become significant for low-energy protons as they approach the Bragg peak. 

\section{Conclusions and outlook}

Newly developed MiniPIX Timepix3 detectors with epitaxially grown 4H-SiC sensors were manufactured, per-pixel energy calibrated and tested with well-defined radiation sources. In this work we evaluated the device spectral-tracking response to the detection of protons with different energies ranging from 13 to 226 MeV. At each energy a rotation scan was performed to obtain the detector response at different angles between the sensor normal and proton beam. Due to the large dead zone above the 65 $\mu$m thick active region (with a 200 V bias), the tilt of the sensor allows a better detection of high-energy charged particles. The resolving power of particle-type classes is limited due to the small active region. The response of the sensor is homogeneous. A small yield of large-amplitude i.e., high-energy small-pixel tracks require further analysis and possible correction of high-energy per-pixel energy response. 
The detected proton tracks are smooth and continuous which demonstrates the proper energy calibration and suitable low-threshold detection. 

In the future, a new sensor design is prepared to prevent voltage shortages at high voltages, where approximately 300 V is required for full 80 $\mu$m sensor depletion (acquired from the C-V measurement as a function of reverse bias voltage and depth of depletion). The major advantage of the SiC is also its great spectral performance without a significant decrease at elevated temperatures up to hundreds of degrees Celsius. Tests to verify the performance of MiniPIX Timepix3 at elevated temperatures are being prepared together with the radiation hardness test, which is expected to be 10${}^3$ times higher compared to commercially available Si. The high radiation hardness allows us to use this detector also for highly damaging neutron detection. In the following work, the absolute detection efficiency for fast neutrons will be analysed from the existing calibrated data and by further calibrations in well-defined neutron fields.

\acknowledgments

Work in Advacam was performed in frame of Contract No. 40001250020/18/NL/GLC/hh from the European Space Agency. Work at STUBA was supported by the Slovak Research and Development Agency grant APVV-18-0273.


\begin{thebibliography}{99}

\bibitem{HEI13} 
E.H.M. Heijne, et al., \emph{Measuring radiation environment in LHC or anywhere else, on your computer screen with Medipix}, \emph{Nucl. Instrum. Meth. A} 699 (2013) 198-204.

\bibitem{BAL18} 
R. Ballabriga, et al., \emph{ASIC developments for radiation imaging applications: The medipix and timepix family}, \emph{Nucl. Instrum. Meth. A} 878 (2018) 10-23.

\bibitem{single_LET} 
P. Stasica, et al., \emph{Single proton LET characterization with Timepix and artificial intelligence for proton therapy treatment planning}, \emph{Physics in Medicine \& Biology}, {68} (2023) 104001.

\bibitem{proton_pencil_beam} 
C. Oancea, C. Granja, et al., \emph{Out-of-field measurements and simulations of a proton pencil beam in wide range of dose rates with Timepix3: Dose rate, flux and LET}, \emph{Physica Medica}, {106} (2023) 102529.

\bibitem{imaging_1} 
V. Olsansky, C. Granja, et al., \emph{Spectral-sensitive proton radiography of thin samples with the pixel detector Timepix3}, \emph{JINST}, {17} (2022) C04016.

\bibitem{imaging_2} 
C. Granja, C. Oancea, et al., \emph{Energy Sensitive Imaging of Focused and Scanning Ion Microbeams with \mbox{$\mu$}m Spatial and \mbox{$\mu$}s Time Resolution}, \emph{EPJ Web Conf.}, {261} (2022) 01007.

\bibitem{radiation_monitor} 
St. Gohl, et al., \emph{A miniaturized radiation monitor for continuous dosimetry and particle identification in space}, \emph{IOP Publishing}, {7} (2022) C01066.

\bibitem{timepix3} 
T. Poikela, et al., \emph{Timepix3: A 65K channel hybrid pixel readout chip with simultaneous ToA/ToT and sparse readout}, \emph{Journal of Instrumentation}, {9} (2014) C05013.

\bibitem{gaas} 
B. Zatko, D. Kubanda, et al., \emph{First tests of Timepix detectors based on semi-insulating GaAs matrix of different pixel size}, \emph{Journal of Instrumentation}, {13} (2018) C02013.

\bibitem{diamond} 
G. Claps, F. Murtas, et al., \emph{Diamondpix: A CVD Diamond Detector With Timepix3 Chip Interface}, \emph{IEEE Transactions on Nuclear Science}, {PP} (2018) 1-1.

\bibitem{b} 
B. Zatko, A. Sagatova, et al., \emph{From a single silicon carbide detector to pixelated structure for radiation imaging camera}, \emph{JINST} {17} (2022) C12005.

\bibitem{sic_temperature} 
N. Gal, et al., \emph{High-resolution alpha-particle detector based on Schottky barrier 4H-SiC detector operated at elevated temperatures up to 500 °C}, \emph{Applied Surface Science}, {635} (2023) 157708.

\bibitem{flash_stray} 
C. Oancea, et al., \emph{Stray radiation produced in FLASH electron beams characterized by the MiniPIX Timepix3 Flex detector}, \emph{JINST} {17} (2022) C01003.

\bibitem{GRA18} 
C. Granja, et al., \emph{Resolving power of pixel detector Timepix for wide-range electron, proton and ion detection}, \emph{Nuclear Instr. Methods A} 908 (2018) 60-71.

\bibitem{tpx2_mixed_field} 
C. Granja, et al., \emph{Spectral and directional sensitive composition characterization of mixed-radiation fields with the miniaturized radiation camera MiniPIX, Timepix2}, \emph{JINST} {17} (2022) C11014.

\bibitem{novak} 
A. Novak, C. Granja, et al., \emph{Spectral tracking of proton beams by the Timepix3 detector with GaAs, CdTe and Si sensors}, \emph{JINST} {18} (2023) C01022.

\bibitem{jakubek} 
J. Jakubek, \emph{Precise energy calibration of pixel detector working in time-over-threshold mode}, \emph{Nucl. Inst. and Meth. A} {633} (2011) S262-S266.

\bibitem{KRI18} 
F. Krizek, et al., \emph{Irradiation setup at the U-120M cyclotron facility}, \emph{Nucl. Inst. and Meth. A} {894} (2018) 87-95.

\bibitem{HOL08} 
T. Holy, et al., \emph{Pattern recognition of tracks induced by individual quanta of ionizing radiation in Medipix2 silicon detector}, \emph{Nucl. Instrum. Meth. A} 591 (2008) 287–290.

\bibitem{dpe} 
Marek L. et al., \emph{Data Processing Engine (DPE): Data Analysis Tool for Particle Tracking and Mixed Radiation Field Characterization with Pixel Detectors Timepix}, submitted to \emph{JINST} (2023).


\bibitem{novel_LET} 
R. Nabha, et al., \emph{A novel method to assess the incident angle and the LET of protons using a compact single-layer Timepix detector}, \emph{Radiation Physics and Chemistry}, {199} (2022) 110349.

\bibitem{CAR21LET} 
C. Granja, C. Oancea, et al., \emph{Wide-range tracking and LET-spectra of energetic light and heavy charged particles}, \emph{Nucl. Instrum. and Meth. A} 988 (2021) 164901.

\bibitem{SOM22} 
M. Sommer, C. Granja, et al., \emph{High-energy per-pixel calibration of Timepix pixel detector with laboratory alpha source}, \emph{Nucl. Instrum. Meth.} A 1022 (2022) 165957.

\end{thebibliography}
\end{document}